# Methods for protein complex prediction and their contributions towards understanding the organization, function and dynamics of complexes

Sriganesh Srihari[1*], Chern Han Yong[2], Ashwini Patil[3], Limsoon Wong[2]

[1] Institute for Molecular Bioscience, The University of Queensland, St. Lucia, Queensland 4072, Australia.
[2] Department of Computer Science, National University of Singapore, Singapore 117417.
[3] Human Genome Centre, The Institute of Medical Science, The University of Tokyo, 4-6-1 Shirokanedai, Minato-ku, Tokyo 108-8639, Japan.
*Correspondence: s.srihari@uq.edu.au

**Abstract**

Complexes of physically interacting proteins constitute fundamental functional units responsible for driving biological processes within cells. A faithful reconstruction of the entire set of complexes is therefore essential to understand the functional organization of cells. In this review, we discuss the key contributions of computational methods developed till date (approximately between 2003 and 2015) for identifying complexes from the network of interacting proteins (PPI network). We evaluate in depth the performance of these methods on PPI datasets from yeast, and highlight challenges faced by these methods, in particular detection of sparse and small or sub- complexes and discerning of overlapping complexes. We describe methods for integrating diverse information including expression profiles and 3D structures of proteins with PPI networks to understand the dynamics of complex formation, for instance, of time-based assembly of complex subunits and formation of fuzzy complexes from intrinsically disordered proteins. Finally, we discuss methods for identifying dysfunctional complexes in human diseases, an application that is proving invaluable to understand disease mechanisms and to discover novel therapeutic targets. We hope this review aptly commemorates a decade of research on computational prediction of complexes and constitutes a valuable reference for further advancements in this exciting area.

# 1. Introduction

Most biological processes within cells are carried out by proteins that physically interact to form stoichiometrically stable *complexes*. Even in the relatively simple model organism *Saccharomyces cerevisiae* (budding yeast), these complexes are composed of several subunits that work in a concerted manner. These complexes interact with individual proteins and other complexes to form functional modules and signalling pathways that drive the cellular machinery. Therefore, a faithful reconstruction of the entire set of complexes is essential not only to understand complex formation but also the higher level functional organisation of cells.

High-throughput experimental systems including yeast two-hybrid (Y2H), tandem affinity purification followed by mass spectrometry (TAP-MS) and protein complementation assay (PCA) have mapped a considerable fraction of interactions from model organisms including *S. cerevisiae* [1-7], *Drosophila melanogaster* [8,9] and *Caenorhabditis elegans* [10], thereby fuelling computational methods to systematically analyse these large-scale interaction data. Beginning from classical methods by Spirin & Mirny [11] and Bader & Hogue [12] that work primarily by clustering the network of protein interactions (PPI network), computational methods have come a long way, and current methods integrate diverse information with PPI networks to predict complexes. These methods have been tested extensively on data from model organisms [13, 14], and are currently being extended to identify and catalogue complexes from less extensively mapped organisms including *Homo sapiens* [15].

Protein complexes represent *modular* functional units within the PPI network [11]. From a biological perspective, this modularity ensures division of labour and provides robustness against mutation and chemical attacks [16]. From a topological perspective, this modularity represents densely connected groups of proteins that function as complexes [17]. Most methods identify complexes by mining modular or dense subnetworks from PPI networks. While this general strategy looks straightforward, these methods are severely restricted by limitations in existing PPI datasets, in particular by the lack of sufficient interactions between "complexed" proteins and the presence of a large number of false-positive (noisy) interactions [18, 19]. Therefore, increasing the interaction coverage by integrating PPI datasets from multiple studies and reducing the noise by assessing the reliabilities of interactions (scoring of PPIs) [20-22] are crucial for accurate complex detection. To summarise, computational identification of complexes from experimental datasets involves the following steps:

  i. Integrating interactions from multiple experiments and assessing the reliabilities of these interactions;
 ii. Constructing a reliable PPI network using only the high-confidence interactions;
iii. Identifying modular subnetworks from the PPI network to generate a candidate list of complexes; and
 iv. Evaluating these candidate complexes against *bona fide* complexes and validating and assigning roles for novel complexes.

Over the last ten years, more than 20 different methods have been proposed in the literature for complex prediction from PPI networks. From time-to-time, surveys have evaluated these methods on datasets available at the time. For example, one of the earliest comprehensive evaluation of methods were by Brohee & van Helden [23] and Vlasbloom *et al.* [24], who compared these methods on yeast Y2H datasets. Subsequently, Li *et al.* [13] and Srihari & Leong [14] presented a more exhaustive evaluation by including raw and scored yeast datasets from TAP-MS and PCA studies [6, 7, 25]. More recently, Yong & Wong [26] studied these methods specifically for the deconvolution of overlapping complexes in dense regions of PPI networks, the recovery of complexes in sparse regions of PPI networks, and the recovery of small complexes in PPI networks. With increasing coverage for human PPI datasets [27-30] these methods are now being applied to predict human complexes [15].

The resources for *bona fide* complexes against which predicted complexes are evaluated have also expanded over the years. For example, the Munich Information Centre for Protein Sequences (MIPS) (http://mips.helmholtz-muenchen.de/proj/ppi/) [31] and the Curated Yeast Complexes (CYC) 2008 (http://wodaklab.org/cyc2008/) [32] databases contain more than 400 experimentally validated complexes for yeast, whereas COmprehensive ResoUrce of Mammalian protein complexes (CORUM) (http://mips.helmholtz-muenchen.de/genre/proj/corum) [33] contains over 2000 validated 'core' mammalian complexes. Predicted complexes that have been subsequently validated have in turn contributed several novel complexes to these catalogues (*e.g.* http://human.med.utoronto.ca/) [15].

The ability to predict complexes from multiple species makes it is possible to examine the reorganization and rewiring of complexes between these species, and thereby estimate the evolutionary conservation of complexes [9]. This could potentially have far-reaching implications, for example, in translating therapeutically relevant observations from model organisms to human [34,35]. For example, Nguyen *et al.* [36] note that rewiring and reorganisation of complexes from yeast to human can affect the transfer of *synthetic lethality* (SL) relationships between genes identified in yeast [37] to human; SL relationships are of therapeutic value in the context of human cancers [38].

Similarly, complexes predicted across disease conditions have revealed extensive rewiring (differential wiring) between these conditions, thereby highlighting key targetable avenues for these diseases [39]. By focusing on rewiring within complexes rather than of the entire PPI network definite dysfunctional regions could be located, thus identifying therapeutically targetable proteins.

Considering the valuable contributions of complex prediction methods, here we put together an extensive survey of methods developed to date (approximately between 2003 and 2015) and evaluate their performance on yeast PPI datasets. We build on earlier surveys [13, 14, 26] so as not to entirely repeat their findings, but discuss challenges faced by methods more lately, in particular detection of sub- or small and sparse complexes and discerning of overlapping complexes. We discuss these methods in the context of evolutionary conservation of complexes between species. By covering methods that integrate diverse information including gene expression and 3D structures of proteins with PPI networks, we discuss the dynamics of complex formation. Finally, we describe methods to identify dysfunctional complexes in human diseases, an application that is proving invaluable to understand disease mechanisms and to discover novel therapeutic targets.

## 2. Review of methods for complex prediction from PPI networks

Although in general, most methods rely on the assumption that protein complexes are embedded as densely connected proteins within the PPI network, these methods vary considerably in their algorithmic strategies and auxiliary biological information employed to identify complexes. Accordingly, these methods have been classified (Table 1) [14] as (i) those based solely on PPI network topology; and (ii) those based on PPI network topology and additional biological insights. By incorporating functional, structural, organizational or temporal information, these methods overcome some of the limitations of experimental datasets, in particular the presence of noise, thereby improving complex prediction. Several of these methods are available as easy-to-run command-line programs or *Cytoscape* [40] plug-ins (Table 1).

To begin, a PPI network is modelled as an undirected graph $G = (V, E)$ where $V$ is the set of proteins and $E = \{(u,v): u, v \in V\}$ is the set of interactions between these proteins. For a protein $v \in V$, the set of neighbours of $v$ is $N(v)$ and the *degree* of $v$ is $deg(v) = |N(v)|$. The *interaction density* of a subgraph $G'(V', E')$ of $G$ is $\frac{2|E'|}{|V'|.(|V'|-1)}$.

### 2.1 Methods based solely on network clustering

Methods based solely on the topology of PPI network look for dense subnetworks or clusters in the network to identify candidate complexes. While some of these methods adopt an agglomerative approach by beginning with singleton or small sets of proteins and growing these sets based on certain cost criteria, some others adopt a partitioning approach by repeatedly breaking down larger clusters into smaller clusters.

#### 2.1.1 Molecular COmplex Detection (MCODE)

MCODE [12] is one of the first computational methods for predicting complexes from PPI networks. MCODE adopts an agglomerative approach that works in three stages: protein (vertex) weighting, complex extraction and an optional post-processing of complexes.

In the first stage, each protein $v$ in the network $G = (V, E)$ is weighted based on the core-clustering density of $v$, which is measured as the clustering coefficient of the highest $k$-core in the neighbourhood of *v*. In the second stage, the protein *s* with the highest clustering density is used to seed a complex. MCODE then recursively moves outward from *s* by including proteins into the complex whose weights are a given percentage (vertex weight parameter) away from that of *s*. This process stops when there are no more proteins to be added to the complex. If there are seed vertices still available, new complexes are seeded and expanded in a similar manner. The optional third stage performs a post-processing by including proteins from the neighbourhood regions of complexes using a "fluff" parameter: neighbouring proteins whose clustering density is higher than this parameter are included into the complexes. The resultant complexes are then scored and ranked based on their weighted densities.

### 2.1.2 Markov Clustering (*MCL*)

MCL [41] is a fast, highly scalable graph clustering method. Applied initially to cluster protein sequences [42], MCL has proved effective for clustering large PPI networks due to its scalability [43,44].

MCL works by simulating random walks (called a *flow*) to extract dense regions from the network. To simulate the flow, MCL iteratively manipulates the adjacency matrix of the network using two operators, *expansion* and *inflation*, that control the spread and thickness of the flow, respectively. Expansion enables the flow to reach all regions of the network, whereas inflation controls the contraction of the flow, making the flow thicker in dense regions and thinner in sparse regions. In each iteration, these parameters increase the probabilities for the random walks within clusters (intra-cluster walks) and decrease the probabilities for the walks between clusters (inter-cluster walks). This process progressively separates out dense regions within the network, ultimately identifying non-overlapping clusters from the network. Since the entire process is executed as matrix operations, MCL is fast and scalable even to large networks.

### 2.1.3 Clustering based on merging Maximal Cliques (*CMC*)

CMC [45] works by repeated merging of maximal cliques extracted from the PPI network. CMC incorporates reliability scores for PPIs and therefore improves on earlier clique-merging methods, including CFinder [46] and Local Clique Merging Algorithm (LCMA) [47], that work only on unscored networks.

CMC begins by enumerating all maximal cliques in the PPI network using the fast search-space pruning-based Cliques algorithm [48]. Each clique *C* is assigned a score which is the weighted interaction density of *C*, given by $\frac{\sum_{u,v \in C} w(u,v)}{|C|.(|C|-1)}$. Cliques are ranked in non-increasing order of their weighted densities. CMC then iteratively merges highly overlapping cliques

based on the extent of their inter-connectivity. The inter-connectivity $I(C_1, C_2)$ between two cliques $C_1$ and $C_2$ is given by:

$$I(C_1, C_2) = \sqrt{\frac{\sum_{u \in (C_1 - C_2)} \sum_{v \in (C_2)} w(u,v)}{|C_1 - C_2|.|C_2|} \cdot \frac{\sum_{u \in (C_2 - C_1)} \sum_{v \in (C_1)} w(u,v)}{|C_2 - C_1|.|C_1|}}$$

If $I(C_1, C_2) \geq T_m$, a merge threshold, then $C_2$ is merged with $C_1$, or $C_2$ is simply removed if it overlaps significantly with $C_1 : |C_1 \cap C_2| / |C_2| \geq T_o$, an overlap threshold. Finally, all merged clusters are ranked by their weighted densities and output as predicted complexes. Since CMC takes into account the weights of interactions, it prioritises more reliable cliques for the merging process while eliminating the less reliable ones, thereby discounting the effects of noise in PPI datasets.

### 2.1.4 Clustering with Overlapping Neighbourhood Expansion (ClusterONE)

ClusterONE [49] works similar to MCODE, by seeding and greedy neighbourhood expansion. ClusterONE first identifies seed proteins and greedily expands them into groups $V$ based on a cohesiveness measure, given by:

$$f(V) = \frac{w^{(in)}(V)}{w^{(in)}(V) + w^{(bound)}(V) + p(V)},$$

where $w^{(in)}(V)$ is the total weight of interactions within $V$, $w^{(bound)}(V)$ is the total weight of interactions connecting $V$ to the rest of the network, and $p(V)$ is a penalty term to model uncertainty in the data due to missing interactions. At each step, new proteins are included into $V$ until $f(V)$ does not increase. $V$ is then denoted as a locally cohesive group. Highly overlapping groups are merged to produce candidate complexes. Since this step allows for overlapping complexes, ClusterONE enhances the performance of MCODE and MCL.

### 2.1.5 Hierarchical Agglomerative Clustering with Overlaps (HACO)

HACO [50] modifies the classical hierarchical agglomerative clustering (HAC) [51] to identify overlapping complexes. The standard HAC algorithm with average linkage [52] maintains a pool of candidate sets to be merged. The distance between two non-overlapping sets $S_1$ and $S_2$ is given by:

$$d(S_1, S_2) = \frac{1}{|S_1||S_2|} \sum_{p \in S_1, q \in S_2} d(p, q),$$

where $d(p, q)$ is the negative of the affinity between proteins $p$ and $q$. In each step of HAC, two non-overlapping sets $S_1$ and $S_2$ with the closest distance are iteratively merged to generate a new set $S_{12}$, while $S_1$ and $S_2$ are removed. The algorithm terminates when there are no remaining sets to merge.

In HACO, the sets $S_1$ and $S_2$ are retained for later use as required, the intuition being that if there is another set $S_3$ whose distance to $S_1$ is only slightly greater than that of $S_2$ then the decision to merge $S_1$ and $S_2$ could be arbitrary and unstable. In this case, HACO produces two merged sets $S_{12}$ and $S_{13}$ by retaining $S_1$ based on a divergence decision: if $S_1$ is considerably different from $S_{12}$ then $S_1$ is retained (in order to generate $S_{13}$), otherwise $S_1$ is removed while keeping $S_{12}$. This procedure results therefore in overlapping complexes.

*2.1.6 Ensemble clustering*

Yong *et al.* [53] developed that an ensemble clustering approach to aggregate clusters generated from multiple clustering algorithms (including MCL, CMC, ClusterOne and HACO) using a majority voting-based scoring. The intuition behind aggregating clusters from different methods is to improve the coverage of complexes while maintaining the quality of the resultant clusters by scoring higher those predicted by multiple methods.

**2.2 Methods based on network clustering combined with biological insights**

Incorporating auxiliary information with the analysis of PPI networks overcomes some of the inherent limitations of PPI datasets, in particular noise, thus enhancing the performance of complex prediction methods.

*2.2.1 Methods incorporating core-attachment structure*

CORE [54], COACH [55], MCL-CAw [56,57] and CACHET [58] look for clusters that adhere to the *core-attachment* organization, noted originally in yeast complexes by Gavin *et al.* [6]. Large-scale pull-down of yeast complexes using TAP-MS in [6] revealed that proteins within complexes are organized as two distinct sets: *cores* that constitute central functional units of complexes, and *attachments* that aid core proteins in their functions. Consequently, by specifically looking for clusters that adhere to this organization, complexes could be identified with better accuracies.

In CORE [54], the probability for two proteins $u$ and $v$ with degrees $d_u$ and $d_v$, respectively, to belong to the same core is determined by the number of common neighbours $|N(d_u) \cap N(d_v)|$ between $u$ and $v$. The probability that $u$ and $v$ have at least $m$ common neighbours participating in $i$ interactions is computed under the null hypothesis that $d_u$ interactions connecting $u$ and $d_v$ interactions connecting $v$ are assigned to random neighbours in the PPI network. This probability is used to arrive at a *p*-value for $u$ and $v$ to belong to the same core, given by:

$$p\_value(u,v) = \Pr(\geq i \text{ interactions and } \geq m \text{ neighbours})$$

$$= \sum_{i \leq j \leq |E|, m \leq k \leq \min\{d_1, d_2\} - j} P_{interact}\left(j\ ||V|, d_1, d_2\right) \cdot P_{common}\left(k\ ||V|, d_1, d_2, j\right),$$

where $P_{interact}$ and $P_{common}$ are computed under the null hypothesis. The *p*-value for (*u*, *v*) is then compared to *p*-values from all pairs involving *u* and *v*, and if (*u*, *v*) is ranked the highest among all these pairs (*i.e.*, (*u*, *v*) has the lowest *p*-value), then (*u*, *v*) is considered to belong to a two-core {*u*, *v*}.

CORE then repeatedly merges cores of sizes two, three and so on until further increase in core size is not possible, to produce the final set of cores. Subsequently, a protein *p* is added as an attachment to a core if *p* interacts with at least half the members of the core, to produce a complex.

COACH [55] works by identifying small dense neighbourhoods around proteins with high degrees in the PPI network. These dense subnetworks are then merged to generate cores. Attachments are added to these cores in a similar way as CORE to produce complexes.

MCL-CAw [56,57], on the other hand, refines clusters produced from MCL [41] by identifying core and attachment sets of proteins within each cluster to build complexes. A set of densely connected proteins within each MCL cluster is designated as a core, and attachment proteins are then included based on their connectivity to this core to produce a complex. MCL-CAw ensures that these attachment proteins can originate from outside the cluster and can be assigned to multiple cores, thus allowing for overlapping complexes.

CACHET [58] is different from the above methods in that it is specialized for reliability-weighted bipartite graphs of bait-prey interactions produced from TAP experiments. TAP uses immobilized baits proteins to capture prey proteins that interact, thus preserving co-complex relationships among these proteins; such relationships are typically lost when the TAP data are converted to pairwise interactions in PPI networks. CACHET first extracts maximal non-overlapping bicliques from the input bipartite graph as cores, and then assembles, in a similar way as CORE, the attachment proteins of these cores.

### 2.2.2  *Methods incorporating functional information*

Proteins within a complex are generally enriched for the same or similar functions. Therefore, combining functional annotations for proteins where available with the topology of PPI networks could improve complex identification. Following on this idea, the Restricted Neighborhood Search Clustering (RNSC) [59], Dense neighbourhood Extraction using Connectivity and conFidence Features (DECAFF) [60] and Protein Complex Prediction (PCP) [61] make use of functional annotations from Gene Ontology [62] to predict complexes.

RNSC [59] employs a cost minimization strategy to partition the PPI network by iteratively moving proteins between clusters until an integer-valued cost function is optimized. To prevent settling into poor local minima, RNSC periodically shuffles the clustering by dispersing the contents of a cluster at random. Finally, RNSC assigns a *p*-value to each of the clusters based on the functional coherence of the constituent proteins, and outputs only the clusters with $p < 0.001$ as the list of complexes. DECAFF [60] follows a clique-identification and merging procedure to identify clusters from the PPI network, and then filters these clusters using functional coherence of the proteins. On the other hand, PCP [61] uses the functional annotations to assign weights to interactions in the network, and uses these weighted interactions to cluster the network based on clique merging to generate complexes.

## 2.3 Comparative assessment of complex detection methods

Here we compare some of the complex prediction methods described above for predicting complexes from the yeast interactome. We obtain PPI data by combining physical interactions from the BioGRID [27], IntAct [63,64] and MINT [75] repositories. These repositories catalogue interactions detected from a multitude of studies, *e.g.* [1,2] (which employ Y2H), [5] (PCA) and [6,7] (TAP-MS). To assess the reliabilities of these interactions detected using different experimental techniques, we compute the reliability for each pair against a common independent criteria; here using similarities between Gene Ontology [62] annotations for these proteins. Specifically, each interaction ($a$, $b$) is weighted using a metric based on the number and type of experiments that detected the interaction, given by:

$$reliability\ weight\ (a,b) = 1 - \prod_{i \in E_{a,b}} (1 - rel_i)^{n_{i,a,b}},$$

where $E_{a,b}$ is the set of experimental technique that detected interaction ($a,b$); $rel_i$ is the estimated reliability of experimental technique $i$ calculated as the fraction of interactions detected by $i$ such that both partners share at least one high-level Cellular Component term from Gene Ontology [62]; and $n_{i,a,b}$ is the number of times that experimental technique $i$ detected interaction ($a,b$). A weighted PPI network was constructed using the top 20000 interactions, covering 3680 proteins (average node degree 10.87).

A predicted complex (or a cluster) $P$ matches a known complex $C$ if the *Jaccard* similarity between $P$ and $C$, Jaccard($P,C$) ≥ 0.5, where:

$$Jaccard(P,C) = \frac{|P \cap C|}{|P \cup C|}.$$

Given the set of reference complexes $\mathbf{C} = \{C_1, C_2, \ldots C_n\}$, the precision, recall, and F-score of a set of predicted clusters $\mathbf{P} = \{P_1, P_2, \ldots P_m\}$ are given by:

$$Precision = \frac{|\{P_i \in \mathbf{P} \mid \exists C_j \in \mathbf{C}, P_i matches\ C_j\}|}{|\mathbf{P}|}$$

$$Recall = \frac{|\{C_i \in \mathbf{C} \mid \exists P_j \in \mathbf{P}, P_j matches\ C_i\}|}{|\mathbf{C}|}$$

$$F = \frac{2 \cdot Precision \cdot Recall}{Precision + Recall}$$

Predicted complexes are scored by their weighted densities and ranked. We calculate the area under the curve (AUC) of the precision-recall curve.

We employ the CYC2008 catalogue [32] (accessed 2012) set as our reference yeast complexes, consisting of 408 complexes. We evaluate the methods only for prediction of large complexes (consisting of at least four proteins), of which there are 149 in CYC2008. This is because practically all methods find it difficult to detect small complexes (consisting of fewer than four proteins) and hence explicitly exclude these complexes from their predictions (*e.g.* see [55]). Besides, the possibility of a predicted complex matching a reference complex that is small purely by chance is relatively high [61], and therefore evaluating the methods becomes challenging (further discussed in Section 2.4.3).

Figure 1 shows the performance of nine methods using precision, recall, F-measure and AUC. We see that methods incorporating biological information achieve higher recall and also generate ranked predictions with higher AUC compared to those based solely on network clustering. Methods that leverage reliability weights (MCL, CMC, ClusterONE, HACO and MCL-CAw) achieve higher recall than those that ignore these weights (MCODE). These results agree with evaluations from earlier studies [13,14]. Finally, ensemble clustering attains the highest recall while maintaining high AUC. The best-performing methods on an average predict about 75% of the complexes.

Figure 2 shows the neighbourhood subnetworks around two example complexes predicted by some of the complex discovery methods. Figure 2a shows the CBF3 complex, which consists of four proteins; these proteins are connected to a number of external proteins outside the complex making it difficult for some methods to recover this complex with high accuracy. CMC and COACH both recover the complex accurately, whereas RNSC recovers only three proteins and ClusterONE includes one extra (noisy) protein into the prediction. Figure 2b shows the mRNA cleavage factor complex consisting of five proteins. Again, these proteins are connected to many external proteins; furthermore, one of the complex proteins, Hrp1p, is not directly connected to the rest of the complex. As a result, none of the methods predict the entire complex accurately: CMC and COACH both predict four of the five proteins, RNSC predicts three and MCL predicts two along with an external protein.

## 2.4 Open challenges in complex detection

The above examples highlight major challenges in complex discovery: many complexes either do not form dense subnetworks or are too small to be recovered accurately.

### 2.4.1 *Detection of sparse complexes*

Existing methods rely on the assumption that complexes are embedded as dense subnetworks within the PPI network and hence adopt density-based clustering for identifying complexes. In an analysis of complexes identifiable from a yeast PPI network, it was noted that only about 65% of complexes with at least four proteins in the network could be identified with Jaccard similarity ≥ 0.50 [66]. The remaining 35% missed complexes did not meet the denseness criteria due to lack of sufficient interactions between member proteins. Even in the well-studied organism yeast, about 30% of the interactome still remains to be mapped of an estimated 25000 – 35000 interactions [67], thus posing a severe challenge to methods that are based on dense subnetworks to identify complexes. To overcome this limitation, [66] proposed to include functional interactions including association between proteins based on functional similarity to enhance the density of complexed regions within PPI networks, and thereby aid existing methods in identifying sparse complexes. Doing so enhanced the performance of MCL, MCL-CAw, CMC and HACO by up to 47% on average across a number of yeast PPI networks.

Supervised Weighting for Composite Networks (SWC) [53] integrates even more data sources including functional association data derived from multiple evidence such as co-occurrence in the literature, to build a composite protein network which fills in the missing interactions within sparse complexes. To reduce the noise introduced into the network, SWC weights the edges using a supervised-learning approach. This improved the performance of most clustering algorithms in yeast and human complex prediction, with sparse complexes benefitting the most.

### 2.4.2  *Discerning overlapping complexes*

Many proteins participate in multiple distinct complexes, resulting in complexes that overlap in the PPI network. These overlapping complexes are frequently highly inter-connected to each other, making it difficult for clustering algorithms to correctly decipher their boundaries [26]; approximately 40% of yeast complexes overlap with at least one other complex.

Some proteins use the same binding surface to interact with multiple partners so that these interactions do not occur simultaneously. Such mutually exclusive interactions can be used to discount simultaneously occurring interactions, which can help to deconvolute overlapping complexes and produce finer clusters in general. For example, Jung *et al.* [68] used structural data of protein binding interfaces to construct a simultaneous PPI network (SPIN) containing only cooperative interactions and exclude mutually exclusive interactions. MCODE and LCMA displayed considerable improvement on SPIN relative to the original PPI network. Ozawa *et al.* [69] used domain-domain interactions (DDIs) to identify conflicting pairs of protein interactions and used these to refine the clusters from MCODE and MCL. The accuracies of these methods improved by at least two-fold. Similarly, Will & Helms [70] integrated PPI networks and DDIs, taking into account the connectivity constraints due to sharing of domains, to identify transcription-factor (TF) complexes in yeast.

Liu *et al.* [71] reasoned that proteins with many neighbours in the PPI network are unlikely to interact with all of them simultaneously. Such proteins, or hubs, were thus removed before clustering, and added back to the generated clusters to which these were highly connected. Furthermore, since a set of interactions can occur simultaneously only if all interacting partners are in the same cellular compartment, the PPI network was decomposed into spatially coherent subnetworks before clustering. This technique improved the performance of MCL, RNSC, IPCA, and CMC, in part because overlapping complexes could be more easily separated and extracted.

Tatsuke & Maruyama [72] observed that the sizes of protein complexes tend to follow a characteristic power-law distribution wherein the majority of complexes are small whereas the larger complexes occur less frequently. This insight was used to randomly partition the PPI network into complexes (clusters) of different sizes using Markov chain Monte-Carlo sampling [73]. Interestingly, this sampling-based approach (PPSampler) could recover several known complexes from CYC2008. The most recent version Repeated Simulated Annealing of Partitions of Proteins (ReSAPP) [74] uses simulated annealing method to optimize the sampling by returning the partition with the highest probability. ReSAPP combines clusters from multiple sampling runs and thereby can also identify overlapping complexes.

### 2.4.3 *Detection of small complexes*

Small complexes (consisting of fewer than four proteins) comprise the majority of complexes in yeast and human, but their prediction is especially susceptible to inaccuracies in the PPI network: missing interactions could easily disconnect a small complex whereas spurious interactions could embed the complex within a larger subnetwork. Topological measures such as interaction density applicable to large complexes are less effective for detecting small complexes – *e.g.* from a network with $n$ proteins there are $O(n^3)$ triplets (with density 1) that could be predicted as three-protein complexes. Furthermore, evaluation measures such as Jaccard match become less effective for evaluating small complexes – *e.g.* a mismatch of only one protein in a three-protein complex renders the prediction inaccurate or less useful despite achieving a Jaccard of 0.50. As a result, most methods fare poorly in detecting small complexes (evaluated in [26,75]) or explicitly exclude small complexes from their predictions (*e.g.* see [55]). Detection of small complexes therefore requires specialized methods.

Yong *et al.* [75] propose one such specialized method called size-specific supervised weighting (SSS). SSS integrates functional associations and literature co-occurrences with PPI data, along with various topological characteristics, using a supervised approach to weight each interaction with its probability of belonging to a small complex. Small complexes are extracted and scored with their cohesiveness-weighted density, which incorporates interactions both within and surrounding each complex. SSS attains better performance in small-complex prediction compared to traditional clustering approaches, deriving about 50% more small complexes at equivalent precision levels.

Ruan *et al.* proposed two methods for predicting size-two and size-three complexes separately [76,77]. Both methods use weights of the interactions around putative small complexes as well as the number of domains in the constituent proteins to derive features for a kernel-based supervised approach. These methods outperform traditional clustering approaches in predicting heterodimeric and heterotrimeric complexes.

Protein sub-complexes can be considered as an interesting special case of overlapping and/or small complexes in which a subset of proteins from a larger complex forms a smaller but by itself a distinct complex. This can be related to cores in which the set of core proteins interact with different sets of attachments to form distinct complexes [6]. Since these sub-complexes overlap with multiple complexes, most general clustering methods either merge all complexes to produce less discernable large clusters. TAP data (*e.g.* [6, 7]) is valuable here because bait-prey pairs from sub-complexes tend appear multiple times as part of (larger) complexes. Zaki and Nora [78] found that CACHET [58], which is specialized to TAP data, was highly effective in identifying these sub-complexes. Based on this idea, these authors developed TRIBAL (TRIad-Based Algorithm) which preserves the multi-edge nature of these bait-prey interactions in TAP data to identify sub-complexes.

**2.5 Detecting evolutionarily conserved complexes**

With rapid increase in the number of resources for human PPIs over the last several years [27 – 30], applying complex prediction methods to identify human complexes has become feasible, and recently a number of studies have attempted to reconstruct complexes from different human tissues and across diseases states (Section 4). Among the interesting observations is that many human complexes are ancient and slowly evolving, with roughly a quarter of the human complexes overlapping with those from lower-order organisms [15]. This has inspired several studies to look at the evolutionary conservation of complexes between human and lower-order organisms. While some of these studies have mainly looked into the evolutionary convergence or divergence of complexes, others have employed these insights to further enhance complex prediction.

Among the seminal works in this direction were by Kelley *et al.* [79] and Sharan *et al.* [80] who constructed *orthology networks* using conserved interactions between species (initially between *S. cerevisiae* and the bacteria *Helicobacter pylori* and later extended to human) based on protein-sequence homology, and clustered these networks to identify conserved complexes between these species. The complexes so-identified were involved in protein translation, DNA-damage response (DDR) and nuclear transport, suggesting that complexes from core cellular processes tend to be evolutionarily conserved.

Van Dam & Snel [81] studied rewiring of protein complexes between yeast and human by mapping PPI networks onto *bona fide* complexes, and concluded that a majority of co-complexed protein pairs retained their interactions from yeast to human, thereby indicating that evolutionary changes in complexes were not due to extensive rewiring of complexed PPIs but instead due to gain or loss of protein subunits from yeast to human. Hirsh and Sharan [82] devised a probabilistic model of protein evolution and employed it to identify conserved complexes between species. Similar to observations by [79,80], these authors found that complexes involved in core cellular processes including pre-mRNA processing, replication, cytoskeleton maintenance and proteasome were highly conserved.

In an interesting work integrating 3D-protein structural information with PPI networks, Marsh *et al*. [83] characterized the evolutionary conservation of 'pathways of assembly' for complexes. The authors observed that evolutionary events optimized complex assembly by simplifying the topologies of complexes, and thereby demonstrated an evolutionary conservation of the assembly order. In particular, gene fusion events reduced the number of assembly steps by at least one, thereby generating fewer intermolecular interfaces in the resultant complex. These events also optimized network topologies by reducing the number of discrete protein interactions, leading to conservation of complexed regions within networks [84].

Nguyen *et al*. [36] integrated protein domain information with PPI networks to construct *domain-interolog networks* and studied conservation of complexes between yeast and human. These authors noted that although several proteins are conserved by sequence similarity between yeast and human (*e.g.* RAD9 and hRAD9), there are many others that did not show any sequence conservation (*e.g.* BRCA1 in human) and yet performed core functions (*e.g.* cell cycle and DDR) that were conserved. These proteins in fact retained conserved functional domains – for example, the BRCT domain present in yeast RAD9 and human hRAD9 is also present in the non-conserved human BRCA1 and 53BP1; all these proteins play vital roles in DDR [85]. Therefore, considering *functional conservation* by integrating domain similarity rather than mere sequence similarity is important to understand conservation patterns of complexes. Based on domain conservation, the authors found that several human complexes had in fact reorganized *via* creation of "mosaic" proteins that accumulated conserved domains from multiple yeast proteins.

Methods that detect coevolution of interacting proteins could also be used to detect complexes – *e.g.* using insights from studies such as [86] on the coevolution of entire protein sequences and specific interaction sites in the context of protein interactions (also see reviews [87,88]). Sets of interacting proteins that coevolve either tend to conserve their interacting domains or adapt to compensatory changes in binding surfaces of partners, thus suggesting evolutionary pressure possibly to conserve specific functions. Therefore, some of these groups of coevolving proteins could potentially constitute conserved complexes.

# 3. Integrating contextual information with PPI networks for predicting dynamic protein complexes

Many, if not all, protein complexes are *dynamic* entities, which assemble at a specific sub-cellular space and time to perform a specific function and disassemble after that. For example, cyclin-CDK complexes involving cyclin-dependent kinases (CDKs) are activated based on the concentration levels of cyclins in a cell-cycle dependent manner [89]. However, due to the lack of specific contextual (temporal and spatial) information in currently available PPI datasets, it is challenging to decipher the dynamics of complexes solely from PPI networks [71]. This limitation severely impacts the performance of computational methods and more critically our understanding of complex organization and function [90].

## 3.1 Identifying temporal complexes

Several methods have looked into novel ways of integrating contextual information with PPI networks to understand the dynamics of complexes. One of the earliest attempts was by Han *et al*. [91] who integrated expression levels of genes with yeast PPI network to study *hub* proteins. Han *et al*. noted two distinct kinds of hubs that are transiently expressed and interact with other proteins to form dynamic modules – *date hubs*, which interact only with singleton or a small set of proteins at any given time, and *party hubs* which simultaneously interact with several proteins. Although initially contested [92] this finding is now widely accepted [93,94], with Komurov and White [95] further extending the concept to include *family hubs* that constitutively express and interact with other (constitutively expressed) proteins to form static modules.

By integrating PPI networks with the expression levels of cell-cycle proteins, de Lichtenberg *et al*. [96] studied the dynamics of complex assembly and disassembly during the yeast cell cycle. Eukaryotic complexes are composed of both constitutively expressed as well as dynamically expressed proteins, which enable them to assemble "just-in-time" to perform functions. Most subunits of complexes are pre-synthesized and pre-assembled whereas the remaining subunits are synthesized only when required, thereby tightly regulating the final complex assembly: by holding off on the last components, cells prevent accidental 'switching on' of complexes at wrong times.

Similarly, by integrating protein expression levels from the yeast cell cycle with cores and attachments within complexes, [97] found that attachments are enriched significantly higher for dynamically expressed proteins compared to cores, whereas the cores are enriched for constitutively expressed proteins. This pattern reflects the "reusability" of cores during complex formation: cores being reused across multiple complexes are maintained constitutively throughout the cell cycle, whereas attachments being required just-in-time are expressed dynamically when required during complex assembly.

Li *et al*. [98] identified temporal complexes by clustering PPI networks constructed using gene expression data from different experimental time points. Using yeast datasets, Li *et al*.

found that about 60% of complexes existed only at one time point (*i.e.* more dynamic) whereas about 24% of complexes existed in more than three time points (*i.e.* more constitutive). By segregating the PPI network based on time-based profiles, dynamic sub-complexes could be separated from larger static clusters, thereby improving overall complex prediction. Similarly, [99] proposed a method Time Smooth Overlapping Complex Detection (TS-OCD) for joint analysis of PPI networks and time-series gene expression profiles to detect dynamic complexes at each time point. Analysis using yeast datasets showed that significantly many complexes could be detected compared to static methods, and in particular their method could identify complexes that share proteins dynamically to perform time-dependent functions.

Goh *et al.* [100] found that miRNAs with widely different expression profiles (*i.e.*, anti-coexpressed) strongly target hub-spokes in PPI networks but yet avoid targeting the same set of hub-spokes. This suggests that anti-coexpressed miRNAs play an important role in controlling the formation of protein complexes that are mutually exclusive. It is tantalizing to speculate on the possibility of inferring mutual exclusivity proteins which are targets of anti-coexpressed miRNAs, and exploiting this information *via* a SPIN-like approach [68] in protein complex prediction.

### 3.2 Integrating structural information with PPI networks

Incorporating information from three-dimensional (3D) structures of interacting proteins can further aid in the identification of protein complexes. Structural information on interacting proteins has been previously used to identify the nature of the interactions [101]. Proteins using the same interaction interface to bind different partner proteins primarily participate in multiple transient interactions as in the case of several kinases. On the other hand, some proteins use multiple interfaces to bind distinct partners and are often seen as members of obligate complexes [101]. With increasing availability of protein structures, it is now possible to annotate PPI networks with known 3D structures or reasonably accurate homology models [102]. Docking is often used to predict an ensemble of possible macromolecular assemblies of proteins usually through the prediction of complementary binding surfaces on partner proteins [103]. Docking can also be used to identify interacting protein pairs through complete cross-docking, where each protein within a set is docked with all other proteins to identify its potential interaction partners [104-106]. Using the information of known interaction interfaces, or predicted binding sites obtained through evolutionary sequence analysis, can improve the accuracy of interaction partner prediction through cross-docking. The prediction of binding affinities of interacting proteins is of great interest not only for assessing the interactions obtained from high-throughput experiments for their reliability, but also for predicting novel interactions between proteins. However, it is difficult to predict binding affinities using docking scores obtained from current scoring algorithms [107]. Docking is further complicated by the conformational changes that proteins undergo as a result of binding to their cognate partners. These conformational changes include backbone flexibility and movements of amino acid side-chains, both of which can addressed by flexible protein-protein docking methods [108]. Several algorithms and automated tools have been developed for this purpose [109-111]. However, flexible docking is much more difficult than rigid docking and docking protein pairs with large conformational changes is still a challenge.

*3.2.1 Flexibility and intrinsic disorder in protein complexes*

The conformational changes that take place in proteins upon binding correspond to their flexibility in the unbound state [112]. Flexibility is important for the formation of large complexes [113] as well as those containing a greater diversity of subunits [114]. Flexibility allows binding over larger distances and in the form of larger binding interfaces without the loss of entropy [115]. Such flexible proteins participate in dynamic complexes and often contain large regions of *intrinsic disorder*. Intrinsic disorder is an extreme form of flexibility in protein structure.

Intrinsically disordered regions in proteins lack stable 3D structure under physiological conditions and can take on an ensemble of conformations (reviewed in [116] and more recently [117]). The high flexibility of disordered regions allows them to reversibly bind to several partner proteins [118]. Indeed, the presence of intrinsic disorder has long been associated with the ability of proteins bind to multiple partner proteins [119,120] allowing them to play an important role in cell signalling and many other aspects of cellular function [117].

*3.2.2 Fuzzy complexes*

There is increasing recognition for the importance of intrinsic disorder in protein complexes [113], also known as fuzzy complexes [121]. Some complexes show static fuzziness where the disordered region in a protein folds into an ordered conformation on binding by undergoing coupled folding upon binding. An example of this is the induced folding of the disordered pKID (phosphorylated kinase-inducible activation domain) of the transcription factor CREB (cAMP-responsive element-binding protein) binding the KIX domain of CBP (CREB-binding protein) to induce transcription of downstream genes [122] (Figure 3a). Similar folding also takes place in the N-terminal disordered region of p53 on binding the E3 ubiquitin ligase MDM2 [120]. Binding of the disordered region may also take place through the selection of a preformed conformer. For example, the KID (kinase inhibitory domain) of p27$^{Kip1}$, a cyclin-dependent inhibitor, has some preformed helical structure that is used to bind cyclin A and subsequently CDK2 to control cell cycle [123] (Figure 3b). While this was previously proposed as an instance of the induced-fit mechanism of binding [124], the role of the preformed helical structure of p27 in effective binding has been recently identified [125].

Proteins with disordered regions also form dynamic fuzzy complexes where the disordered region stays disordered either partially or completely on binding [126]. Thus, the disordered regions may remain flexible during binding without folding into a fixed structure as in the case of the inhibitor I-2 when it binds PP1 (protein phosphatase 1) [127] (Figure 3c).

Disordered linkers between two ordered domains within a protein also form parts of dynamic fuzzy complexes. They are advantageous because they allow the two domains to sample a large number of orientations with respect to each other, as observed in the calcium-binding domains of calmodulin which adopt different relative orientations when binding to different proteins [128-130].

It has been proposed that in some dynamic complexes, the disordered partner does not bind the ordered partner in a single location but rapidly changes between several conformers binding with the help of a mean electrostatic field rather than through discrete charges [128]. This binding is further affected by post-translational modifications which can change the mean charge presented by the disordered binding interface [128].

*3.2.3 Binding interface and complex prediction*

Intrinsically disordered regions in proteins frequently bind their interaction partners through the use of short linear motifs [129] which adopt different structures when binding different target proteins. On the other hand, proteins may also use molecular recognition features (MoRFs) for binding their cognate partners [130]. MoRFs are short linear regions within disordered segments that participate in specific target recognition and undergo a disorder-to-order transition on binding. Given the flexibility of the binding partners in such complexes, it is difficult to predict their binding sites or protein assemblies. However, several tools have become available in recent years to predict the binding interfaces within the disordered regions. One such tool, SLiMPrints, predicts short linear motifs based on conserved regions that may participate in binding [131]. Other methods use machine learning techniques to identify binding interfaces using a host of features from the sequence of the disordered region including the amino acid propensity of known MoRFs and their flanking regions, physicochemical properties of the amino acids as well as evolutionary profiles [132-135]. It is now possible to use information from experimental techniques like Nuclear Magnetic Resonance (NMR) and Small Angle X-ray Scattering (SAXS) in combination with computational methods to model the ensemble of conformations that may be adopted by disordered regions within protein complexes [114,115]. The Protein Ensemble Database is a collection of such structural ensembles of proteins obtained from a combination of experimental and computational methods [136].

## 4. Identifying complexes in human diseases

Complexes are responsible for driving important mechanisms that maintain cellular homeostasis, but are also often the sites of dysregulation in diseases. The functional analysis of genes within complexes suggests that these complexes could be hotspots for perturbations due to genetic or environmental factors, thereby driving common and rare diseases [137,138].

Identifying complexes dysregulated in human diseases therefore forms an important extension of complex detection methods.

Vanunu *et al.* [139] employed a PPI network to associate complexes with diseases catalogued in the Online Mendelian Inheritance in Man (OMIM) database [140]. Using disease-to-gene associations from the OMIM database as a prior, the proposed method measures the association computed *via* network-propagation between causal genes and genetic diseases. The score computed for each gene is then used in combination with the PPI network to identify complexes involved in the disease. About 566 complexes were identified that were associated with hereditary or congenital diseases including ataxia-telangiectasia (AT), hereditary prostate cancer and microcephalic osteodysplastic primordial dwarfism (MOPD).

Similarly, Lage *et al.* [141] identified about 506 complexes that included disease-promoting genes implicated in disorders such as retinitis pigmentosa, epithelial ovarian cancer, inflammatory bowel disease, amyotrophic lateral sclerosis, Alzheimer disease, type-2 diabetes and coronary heart disease. To do this, the authors constructed a phenome-interactome network to identify candidate complexes, which were scored based on member proteins involved in these disorders, thus prioritising disease-associated complexes from the network.

In diseases such as cancer, cellular dysregulation often involves complexes that regulate critical functions including genome-stability maintenance, cell-cycle checkpointing and control, growth signalling and metabolism, the disruptions of which lead to increased accumulation of genomic instability, cell proliferation and metabolic dysfunction. For example, dysregulation of the BRCA1-A, -B and -C complexes due to loss-of-function mutations or epigenetic silencing of the *BRCA1* gene results in decreased DNA double-strand break repair efficiency, thus contributing to genomic instability in breast cancer [142,143].

Recently, a number of works have attempted to identify complexes dysregulated in cancer by studying rewiring of complexes between cancer conditions. For example, Srihari *et al.* [39, 144] constructed condition-specific PPI networks by integrating gene-expression profiles with human PPI network across normal and cancer conditions in breast and pancreatic cancers. Subsequently, complexes were identified using CMC [45] from each of these condition-specific networks separately and matched to detect rewiring or changes to protein composition within complexes between these conditions. Interestingly, several known cancer genes were involved in these rewiring events, and the affected complexes displayed significant differences in expression levels between these cancer conditions. Among the dysfunctional complexes were those involved in DNA-damage repair (*e.g.* BRCA1 complexes) and growth-factor signalling (*e.g.* EGFR signalling) and also proteasomes, signalosomes and ribosomal complexes.

Zhao *et al.* [145] estimated the differential abundance of protein complexes between normal and cancer conditions in the cancers of 39 human tissues by using gene expression profiles. The authors employed known human complexes from the CORUM database [33], and estimated complex abundance by computing the optimal number of proteins required to form each complex and the number of copies of proteins present in the cell. Complexes involved in DNA-damage repair (*e.g.* BLM-TOP3A), histone modification (*e.g.* HDAC complexes), minichromosome maintenance (*e.g.* MCM complexes) and protein translation (*e.g.* RNA polymerases) showed abnormal expression in several human cancers.

Chen *et al*. [146] modelled disease-complex prioritisation in a network as an optimisation problem by maximizing the information flow between a query disease and a candidate complex through the network. For a queried disease, this approach identified the associated proteins and the complexes formed by these proteins in the network. Application of this approach to breast cancer yielded six complexes involved in DNA-damage repair (BRCA1 and MSH complexes), replication factor (RFC complexes) and chromatin remodelling (SWI/SNF complex).

Goh *et al.* [147,148] demonstrated that analysing proteomic profiles in the context of protein complexes significantly improved the reproducibility and sensibility of biomarker identification from proteomic data. They introduced the concept of proteomic signature profile (PSP), which is a vector of protein complexes and their "hit rates" (*i.e.*, the proportion of protein components detected in a patient sample for the respective complexes) irrespective of individual protein's quantitation level. Complexes that were significantly differential in their hit rates between cases and controls were reported. They uncovered in liver cancer, an interesting relationship between the purine metabolism pathway and two other complexes involved in DNA-damage repair, suggesting progression to poor-stage liver cancer requires additional mutations that disrupt DNA-damage repair enzymes.

## 5. Conclusion

With increasing availability of PPI and other functional datasets, prediction of complexes has come a long way over the last several years. Apart from widely studied model organisms such as yeast [1-7], fruit fly [8, 9] and worm [10], it is now possible to predict complexes from more sophisticated organisms including human [15]. This has provided new opportunities to study complexes under different contexts and across species, thus tracing the functional and evolutionary conservation of complexes. Integrating diverse information including 3D structure and time or context-based gene-expression profiles has helped to map the dynamics of complex formation and also understand their roles in diseases. This forms an excellent example where a fundamental problem such as complex prediction has had far-reaching applications in understanding the organization and functions of the cell. We hope that this review aptly commemorates these efforts and inspires further advancement of research in this exciting area.

**Acknowledgement**

We thank Dr Alison Anderson (The University of Queensland) for critical reading of the manuscript. SS is supported by an Australian National Health and Medical Research Council (NHMRC) grant# 1028742.

**Figures**

**Figure 1: Comparative assessment of complex-prediction methods on yeast datasets.** Ensemble refers to ensemble clustering (combining MCL, CMC ClusterONE, IPCA, COACH and RNSC) [53].

**Figure 2: Examples of complexes predicted by different methods.** Blue nodes are proteins within the complex and red nodes are proteins not in the complex. (a) CBF3 complex, consisting of four proteins, compared with that predicted by CMC and COACH (blue), RNSC (red), ClusterONE (green). (b) mRNA cleavage factor complex, consisting of five proteins, compared with that predicted by CMC and COACH (blue), RNSC (green), MCL (red).

**Figure 3: Flexibility and intrinsic disorder in protein complexes.** (a) Induced folding of the intrinsically disordered pKID domain of CREB (orange) on binding to the KIX domain of CBP (blue) (PDB id: 1KDX). (b) Binding of the intrinsically disordered KID domain (orange) of p27$^{Kip1}$ to the Cyclin A (grey) and Cdk2 (blue) (PDB id: 1JSU). (c) I-2 (orange) stays disordered when binding to PP1 (blue). Disordered regions are indicated by dotted lines and not visible in the X-ray crystal structure (PDB id: 2O8G).

**Table 1: Methods for protein complex prediction from protein interaction networks.** Associated softwares are available as *Cytoscape* [40] plug-ins (Cy), command line programs (CL) or as online (OL) web servers under the mentioned links.

| Classification | Method | Availability (URL) | Reference |
|---|---|---|---|
| *Solely network clustering* | MCODE (Cy) | http://apps.cytoscape.org/apps/mcode | [12] |
| | MCL (Cy, CL) | http://micans.org/mcl/ <br> http://apps.cytoscape.org/apps/clustermaker | [41-43] |
| | CMC (CL) | https://www.comp.nus.edu.sg/~wongls/projects/complexprediction/CMC-26may09/ | [45] |
| | ClusterONE (Cy) | http://apps.cytoscape.org/apps/clusterone | [49] |
| | HACO (CL) | http://www.bio.ifi.lmu.de/Complexes/ProCope/ | [44,50] |
| | PPSampler (CL) | http://imi.kyushu-u.ac.jp/~om/PPSamplerVer1.2/PPSamplerVer1_2.exe | [72] |
| *Core-attachment structure* | CORE (CL) | http://alse.cs.hku.hk/complexes/ | [54] |
| | COACH (CL) | http://www1.i2r.a-star.edu.sg/~xlli/coach.zip | [55] |
| | MCL-CAw (CL) | https://sites.google.com/site/mclcaw/ | [56,57] |
| *Functional information* | RNSC (CL) | http://www.cs.utoronto.ca/~juris/data/ppi04/ | [59] |
| | PCP (CL) | https://www.comp.nus.edu.sg/~wongls/projects/complexprediction/PCP-3aug07/ | [61] |
| *Evolutionary information* | NetworkBLAST (OL) | http://www.cs.tau.ac.il/~bnet/networkblast.htm | [79,149] |
| | NetworkBLAST-M (CL) | http://www.cs.tau.ac.il/~bnet/License-nbm.htm | [79,149] |
| | COCIN (CL) | https://sites.google.com/site/cocinhy/ | [36] |
| *Mutual exclusive interactions* | SPIN (CL) | https://code.google.com/p/simultaneous-pin/ | [68] |
| | DACO (CL) | http://sourceforge.net/projects/dacoalgorithm/ | [70] |
| *Sparse complexes* | SWC (CL) | http://www.comp.nus.edu.sg/~wongls/projects/complexprediction/SWC-31oct14 | [53] |

| *Small complexes* | SSS (CL) | http://www.comp.nus.edu.sg/~wongls/projects/complexprediction/sss-3dec2014.zip | [75] |
|---|---|---|---|
| *Temporal complexes* | TS-OCD (CL) | http://mail.sysu.edu.cn/home/stsddq@mail.sysu.edu.cn/dai/others/TSOCD.zip | [99] |
| *Complexes in diseases* | CONTOUR (CL) | https://sites.google.com/site/contourv1/ | [39] |